\documentstyle[twoside,fleqn,espcrc2]{article}

\newcommand{\AmS}{{\protect\the\textfont2
  A\kern-.1667em\lower.5ex\hbox{M}\kern-.125emS}}

\title{Some Examples in the Realization of Symmetry \thanks{Talk given
at Strings '97, 18-21 June 1997, Amsterdam, The Netherlands.  To appear
in Nuclear Physics Proceedings Supplement.}}

\author{Frank Wilczek\address{Institute for Advanced Study, 
        School of Natural Science \\ 
        Olden Lane, Princeton, NJ 08540}
        \thanks{Research supported in part by DOE grant DE-FG02-90ER40542.}}
       
\begin{document}

\begin{abstract}
I briefly discuss three phenomena arising in recent work where 
the realization of
symmetries in quantum mechanics is unusual.  The first of these is, I
believe, the very simplest realization of  non-trivial
confinement by a mechanism of charge-flux frustration.  It arises in models
having several coupled abelian Chern-Simons gauge symmetries, which
closely resemble effective theories used for the quantum
Hall effect.  The second is symmetry obstruction by non-abelian flux,
which has implications for 2+1 dimensional supergravity.  Third is the
possibility of new varieties of quantum statistics: non-abelian
and projective.   
\end{abstract}

\maketitle

\section{Apology}

I don't think I have much to teach this audience about string
theory as such.  What I will do, that I hope is useful, is discuss
a few phenomena that have come up in my recent work which are in some way
connected to fundamental issues regarding the realization of symmetry
in quantum mechanics.  My strong impression is that the fundamental
symmetry of string theory, or whatever string theory 
is evolving into, has yet to
be fully elucidated; in trying to do so, it may be well to have
the possibility of unusual realizations in mind.  Also, the examples I
will be considering 
involve in simple low-dimensional gauge field
theories of a kind very similar to those often arising in world-sheet or
membrane studies.  

\section{Simple Models of Confinement}

A classic mechanism of confinement is abstracted from the behavior of
type II superconductors\cite{super}.  These materials confine
magnetic flux in tubes, so that they would induce a linear confining
potential between magnetic monopoles.  It has long been conjectured
that an analogous mechanism could underly confinement in 3+1
dimensional relativistic quantum field theories\cite{mandel}, and
recent marvelous work has supported this idea, at least in the
supersymmetric case\cite{shifman}.  There are, however, several
significant limitations to the analogy: perhaps most important, the
confinement applies to a strictly additive quantum number, and does
not exhibit saturation, as one has for triality; also, the monopole is
not an elementary object, and it is not in general clear that the
degrees of freedom needed to construct it are available or that it is
a plausible stand-in for quarks; finally, related to this, the
detailed implementation tends to be complicated and indirect.  Now I
would like to show you a class of field theories that exhibit
confinement in a very simple and transparent way through charge-flux
frustration\cite{cornalba}, and where the elementary quanta are
confined and the confinement saturates.  Considered as models for QCD
these field theories have weaknesses of their own: they live in 2+1
dimensions, the dynamics seems to be essentially abelian, and the
confinement is only logarithmic -- but at least they are different
weaknesses.  These theories are closely related to the effective
theories used to describe hierarchical and layered states in the
quantum Hall effect.  From a conceptual standpoint, it is particularly
interesting that they contain no massless particles in their
perturbative spectrum; thus they exhibit, in a very clear-cut way,
confinement arising as a global phenomenon, rather than through
accumulation of infrared singularities.

The  Lagrangians  of interest to us take the form
\begin{equation}
{\cal L} ~=~ 
{1\over 8\pi } \epsilon^{\alpha \beta \gamma}  
\mu^{lm}  a^l_\alpha f^m_{\beta\gamma} 
+ D_\alpha \phi^r  (D^\alpha \phi^r )^*  
- V(\phi ) ~, 
\end{equation}
where the covariant derivative on the $r^{\rm th}$ scalar field acts as 
$
D_\alpha \phi^r \equiv \partial_\alpha \phi^r - i q^r_l a^l_\alpha \phi^r
$
(suspending the summation convention on $r$),
and $V$ is an effective potential whose details need not concern
us.  $\mu$ is a symmetric matrix.  The normalization is chosen in
such a way that integral entries in $\mu$ and integral charges arise
naturally, for example if the theory results from breakdown of an
overlying nonabelian theory.

Let us first consider the simple case when $\mu$  is a 2$\times$2 matrix
whose only
non-zero entries are equal to an integer $n$ off the diagonal.  Assume that
just one scalar field, carrying charges $(0, q)$ with respect to the
two gauge groups, condenses.  The field
equations of most interest to us arise from the variation of $\cal L$
with respect to the time-component of $a^1$ and with respect to the
space-components of $a^2$.  Allowing for an external source carrying 
charge of the first type, these equations read:
\begin{equation}
{n\over 2\pi } b^2 = \rho^1_{\rm ext.} 
\end{equation}
and 
\begin{equation}
{n\over 2\pi } {\vec e}^1 = \hat z \times {\vec j}^2 ~. 
\end{equation}
Let us analyze the second of these first. 
If we require that the condensate lives on the vacuum manifold at
infinity, and is single-valued, then by a gauge transformation it
can be brought into the form $\phi \rightarrow v e^{il\theta}$,
where $l$
is an integer and $v$ is magnitude of the vacuum expectation value of
$\phi$.  
In order that the current
due to the condensate
fall off faster than $1/r$ at infinity we must further require that
\begin{equation}
l ~=~ q a^2_\theta ~, 
\end{equation}
in order that the azimuthal covariant derivative vanish.  
Using
Stokes' theorem,
the second of these integrates to the flux quantization condition
$
2 \pi l ~=~ q \Phi^2 ~.
$
Integrating the first equation over all space, on the other hand, gives us
$
Q^1_{\rm ext.}  ~=~ {n\over 2\pi } \Phi^2~.
$

Combining these, we find the condition
\begin{equation}
Q^1_{\rm ext.} ~=~ {n l\over q}~.
\end{equation}
on the external charge.  If $n$ does not divide $q$, this condition
will generally fail, even for integer $Q^1_{\rm ext.}$.  In that case,
it will not be possible for the fields to fall off faster than $1/r$
at infinity.  The resulting long-range fields lead to energy diverging
logarithmically with distance, proportional to the square of the
fractional part of ${q\over n}Q^1_{\rm ext.}$.  Energy of this sort
will arise from the gradient energy of the condensate, or (if a
Maxwell term is present in an expanded form of $\cal L$) the Coulomb
field energy, or both.  Thus charges which do not satisfy the
condition\cite{split} are confined.  On the other hand, from the form
of the condition it is clear that an appropriate multiple of a
confined charge will not be confined.  Thus finite-energy states will
include baryon-like objects, but not the corresponding quarks.

In the more general 2$\times$2 case, when  
$
\mu ~=~ \left(\matrix{m_1 &n \cr n& m_2\cr}\right)
$
and the condensate has charge vector $(q^1, q^2)$,  by following steps similar
to those above, we find conditions
\begin{equation}
Q^1_{\rm ext.} ~=~ {1\over 2\pi} ( m_1 \Phi^1 + n \Phi^2 ) + q^1
\lambda 
\end{equation}
\begin{equation}
Q^2_{\rm ext.} ~=~ {1\over 2\pi}  (n \Phi^1 + m_2 \Phi^2) + q^2
\lambda 
\end{equation}
and the flux quantization condition
\begin{equation}
2\pi l ~=~ q^1 \Phi^1 + q^2 \Phi^2~,
\end{equation}
for the long-ranged fields to vanish.  
Here $\lambda$ is a continuous parameter, representing the ability of
the condensate to screen electric charge.  The ratio of screening 
charges, of
course, must follow that of the condensate.  Now in the generic case,
after solving the flux quantization condition, 
there will be two continuous parameters
$\lambda$ and $\Phi^2$ available to satisfy the remaining two
equations, 
and then an arbitrary charge will be screened (not
confined). However if
$
m_2(q^1)^2 - 2n (q^1)(q^2) + m_1 (q^2)^2 ~=~ 0 ~,
$
then these two parameters multiply proportional coefficients. 
Hence there will be charges that cannot be screened, and
must be confined.  (Our earlier case corresponds to $m_1 = m_2 = q^1 =
0$.)  A brief calculation reveals that the condition 
$Q\mu^{-1}q = l$ which must be
satisfied by unconfined charges can be written 
in the transparent form 
\begin{equation}
q^1Q^2_{\rm ext.} - q^2Q^1_{\rm ext.} = \sqrt{-\Delta} l~, 
\end{equation}
where the determinant $\Delta \equiv m_1m_2 -n^2$.
{}There will be non-trivial real solutions $(q^1, q^2)$ 
if and only if $\Delta \leq 0$;    
there will be integer solutions if and only if $-\Delta$ is a
perfect square.  

It is amusing to see simple number-theoretic conditions emerging in
confinement criteria.  Similar but more intricate phenomena
occur for larger numbers of
species.  They have not been explored systematically.

\section{Symmetry and Supersymmetry Obstruction\cite{split}}

A nonabelian vortex is characterized in regular gauge
by a non-zero matrix vector potential
$a_\theta$ at spatial infinity, and the holonomy
$h ~=~ {\cal P} e^{{i\int^{2\pi}_0} a_\theta~ d\theta}~.$
Here ${\cal P}$ denotes path ordering, and $a$ is a matrix in the
adjoint representation.  (More precisely, $h$ is defined only up to
conjugacy, since a gauge transformation $\Lambda(r, \theta )
\rightarrow \Lambda_\infty ( \theta)$ takes 
$h \rightarrow \Lambda_\infty (2\pi ) h \Lambda_\infty (0)^{-1}$.)  
We may suppose, after a non-singular gauge
transformation, that 
$a_\theta$ is a constant matrix.  Given a charged field $\eta$, one
has its covariant derivative
$
D_\mu \eta ~=~ \partial_\mu \eta + i {\rm T}^a a^{\rm a}_\mu \eta~.
$
Here ${\rm T}^{\rm a}$ are the generators for the representation to which 
$\eta$ belongs, and of course the $a^{\rm a}$ are the internal
components of $a$.  If one diagonalizes ${\rm T}^{\rm a} a^{\rm a}_\mu$
(assumed constant) and considers the partial wave
$\eta (r, \theta ) = e^{il\theta} f(r)$, then (in the regular gauge) $l$ is
quantized to be an integer, but for the $k^{\rm th}$ eigenvalue
$e_{(k)}$ and eigenvector $\eta_{(k)}$ one has the azimuthal
derivative
$
D_\theta \eta_{(k)} ~=~ i(l + e_{(k)} ) \eta_{(k)} ~.
$
And so the angular energy term, which has direct physical significance, 
is proportional to ${(l + e_{(k)})^2\over
r^2}$.  It behaves as if there is a fractional contribution $e_{(k)}$ 
to the
angular momentum.  
Alternatively, by a singular gauge transformation
(with gauge function proportional to $\theta$, and thus having a cut)
one could formally remove $a_\theta$, at the cost of introducing
boundary conditions of the form
$
\eta (2\pi ) ~=~ \rho (h) \eta (0) 
$
on the azimuthal dependence of the field $\eta$ (and thus on the wave
functions of its quanta).  Here of course $\rho$ is the appropriate
unitary representation matrix.  In this formulation the azimuthal
covariant derivative is simply the ordinary spatial azimuthal
derivative, but one has the condition 
$
l_{(k)} ~=~ {\rm integer}~ +~ {\phi_{(k)}\over 2 \pi } 
$ 
on the partial waves $\eta_{(k)} \propto e^{il\theta} f(r)$.  The form
of the angular energy term, of course, is the same as before, with $e_{(k)}
\equiv \phi_{(k)}/2\pi$.  

Generically states with a given value of $e_{(k)}$ (and the same
spatial wave functions) will be degenerate, but states with different
values of $e_{(k)}$ will not be degenerate, because they see a different
effective Hamiltonian -- or, alternatively, because they
obey a different angular
momentum quantization condition.
Now suppose that the symmetry of the 
ground state is a non-abelian group $K$, and
consider $\kappa ~\in~ K$.  If $\rho$ is a faithful representation, and 
$h\kappa \neq \kappa h$, then $\rho (\kappa )$ will not act within the
spaces of fixed $e_{(k)}$, but will connect states with different
values of $e_{(k)}$.   As we have seen, generically this means that
$\rho (\kappa )$ will connect states of different energy.  Thus the
obstructed symmetry $K$ will not be realized as a symmetry of the
spectrum, but will connect states with different energy.  Only 
members of the 
centralizer group ${\cal C} (h)$ of transformations which commute with $h$
will generate spectral symmetries.

The effects of obstructed symmetry appear directly only in states
carrying both flux and charge.  These can be brought together in a
straightforward way, as bound states, with the flux realized as a
classical field.    A more intrinsic form of
the phenomenon occurs in Chern-Simons theories, where the interaction
itself 
naturally associates flux with charge.  The question then arises
whether splittings are induced for the fundamental quanta, which carry
both charge and flux, and what is the mechanism.  
I expect a splitting does arise, for the
following reason.  In evaluating the self-energy of such a quantum,
one must sum over space-time trajectories where its world line is
self-linked or knotted.  The charge from one part of the trajectory gets
entangled with the flux emanating from another part, and the
propagation is altered, as above.  For non-dynamical sources
(Wilson lines) it is known that the amplitude of a trajectory depends
on its topology; indeed, this fact lies at the root of the
application of non-abelian Chern-Simons theories to topology.  It
would be very interesting to do a dynamical calculation along these lines.

In 2+1 dimensional gravity 
the primary effect of a point mass is to
create a conical geometry in its exterior, with a deficit angle $\delta$
proportional to the mass; $\delta = 2\pi Gm$.  
The exterior geometry is then locally flat,
but non-trivial globally.  Indeed, it induces a modification in the
angular momentum quantization for particles in the exterior, similar
in some ways to the non-abelian vortex.  A notable difference is that
whereas the non-abelian vortex induced an additive change in the
quantization condition, change induced by 
the gravitational field of a point mass
is multiplicative. Indeed, from requiring that the azimuthal factor
$e^{il\theta}$ be single-valued as 
$\theta \rightarrow \theta + (2\pi - \delta)$ 
we find $l = {\rm integer}/(1  - {\delta\over 2\pi} )$.  
This difference reflects the
different symmetry of the sources: an additive shift in angular
momentum requires an intrinsic orientation violating the discrete
symmetries P and T, while a multiplicative factor is even under these
transformations\cite{jackiw}.   

The conical geometry presents a global obstruction to defining a
spinor supercharge (see below), so that supersymmetry, even if valid
for the ground state, is obstructed for massive states.  This fact
plays a role in some speculative ideas for using supersymmetry to zero
the cosmological term, while remaining consistent with its absence in
the physical spectrum\cite{witten,becker}.  One can reasonably expect
that fermion-boson degeneracy is lifted, but a concrete calculation
might be welcome.  It is not altogether trivial to produce one,
because the expected effect comes from a subtle infrared phenomenon,
so that one needs to do some sort of non-perturbative calculation; but
on the other hand the theory is non-renormalizable.

One can remove the effect of the conical geometry by a singular gauge
transformation, in favor of a modified boundary condition.  Here the
covariant derivative should respect parallel transport, so that we
should require that the effect of orbiting the apex of the cone is to
rotate spinors (or vectors) through the appropriate angle, {\it
i.e}. to multiply them by a phase proportional to the spin and angle.
An important subtlety, which arises even in flat space, 
is that one must include a factor -1 for half-odd integer (fermion) fields. 
This factor is essentially equivalent 
to the -1
accompanying fermion loops in Feynman graphs, and its inclusion
implements the normal spin-statistics relation.  With this factor 
$e^{2\pi is}$ taken out, the appropriate condition on wave functions
$\psi_s \propto f(r) e^{il \theta}$ for spin $s$ quanta, 
in order that they transform properly, is 
$
\psi_s (2\pi - \delta) ~=~ e^{i (2\pi - \delta )l - i \delta s } \psi_s (0) 
$
leading to 
\begin{equation}
l ~=~ {1\over 1- {\delta\over 2\pi}} {\rm integer} ~+~ 
 s {{\delta\over 2\pi}\over 1 - {\delta\over 2\pi}} ~.
\end{equation}

This superelectrodynamics can be coupled to
supergravity, 
and one can consider gravitational corrections to the
`hydrogen' spectrum. 
One can, in this context, consider the corrections to the effective 
non-relativistic Hamiltonian due to one photino exchange. The computation 
closely follows the classical derivation of 
the Breit Hamiltonian for positronium and yields the correction
\begin{equation}
\Delta {\cal H} ~=~ {-i\over 2m } \bigl [ P_i, V(x)\bigr ] \Sigma_i~,
\end{equation}
where $m$ is the electron mass, $V$ is the Coulomb  potential ${e^2\over 
2\pi} \ln r $, and the $\Sigma_i$ are matrices acting in the internal spin 
space of the electron-proton system.   
The form of this correction to the 
Hamiltonian is not altered by gravity, since $\Delta{\cal 
H}$ is a local operator on the wavefunction and gravity in $2+1$ 
dimensions has only global, topological effects.  This is 
consistent with the boundary conditions
mentioned above; indeed the exact form of the matrices 
$\Sigma_i$ implies that, in a conical geometry, the operator 
$\Delta{\cal H}$ becomes hermitian, in a non-trivial way.
The spin 
matrices connect components of the wavefunction with internal spin 
differing by $1$, and the phase acquired by the operator $[ P_i, 
V(x)\bigr ]$ (or simply by $x_i$ after taking the commutator), which
would otherwise spoil hermiticity, is exactly 
compensated by the different, spin-dependent, boundary conditions on
these components.

The main effect of the gravitational field, 
however, arises already in the nominally
spin-independent interaction, which dominates in the low-velocity
(non-{\it special\/} relativistic) limit: the modified condition
for the allowed angular momenta modifies the effective
Schroedinger equation, and splits the spectrum corresponding to
different spin values.  Here we are concerned with the motion of the
electron in the geometry and potential provided by the proton, and take
$\delta = 2\pi GM$.  Clearly supersymmetry is obstructed,
and it is not manifest in the spectrum.  

We have glossed over several technicalities that do not substantially
affect this leading-order calculation, though 
in extending it to higher orders in $e^2$ and 
$m/M$ they would need
careful attention.  The
most interesting, perhaps, comes from the double-conical geometry
arising when one includes the gravitational fields of both particles.
One then obtains a modified quantization condition on the relative
angular momentum, which takes the form
$
l_{\rm rel.} ~=~ {{\rm integer}\over 1 - {\delta_{\rm max}\over 2\pi}}
~+~ 
{1\over 2\pi }{(\delta_1 s_2 + \delta_2 s_1)   
\over 1 - {\delta_{\rm max}\over 2\pi}}~,
$
where $\delta_i, s_i$ are the deficit angle and spin associated
with particle $i$, and $\delta_{\rm max}$ is the larger deficit angle.  
Also one should use the Kerr geometry to take into account the
spin and angular momentum of the particles, but these effects 
are 
subleading, of relative order $(s~{\rm or}~ l)\delta$.

\section{Nonabelian and Projective Quantum Statistics}

Exotic forms of quantum statistics, apart from the familiar examples
of boson and fermion, are a logical possibility inherent in the
quantum kinematics of identical particles\cite{wilczek}.  In general,
the issue is how to add the amplitudes for processes described by
topologically distinct classes of trajectories.  The classical
Lagrangian, which is designed to yield the equations of motion at its
extrema, does not necessarily answer this question, and one can
imagine a range of models having a single classical limit but
fundamentally different quantum behavior.  This occurs for QCD, in
connection with the $\theta$ parameter.  There it is the topology of
trajectories in field space which is relevant; but the same question
arises even for trajectories of identical point-like particles, and
this is the case we associate with quantum statistics.

This problem takes a special form in low dimensions.  In 2+1
dimensions the topological classes of trajectories are parametrized by
the braid group, and one can classify the possibilities for quantum
statistics by studying the unitary -- or, in principle, projective --
representations of that group.  The one-dimensional representations of
the braid group describe anyons.  The quasiparticles and quasiholes of
the conventional Laughlin and `hierarchical' fractional quantum Hall
states are anyons, and not simply bosons or fermions.

It is natural, in this context, to ask if nonabelian representations
of the braid group also occur in physics.  Their implications are quite
unusual.  Since their irreducible representations are
multi-dimensional, nonabelian representations of quantum statistics
require a degeneracy in Hilbert space for the particles at fixed
positions in physical space, with adiabatic transport of the particles
around one another (in the absence of any explicit interaction) giving
not merely a phase factor, but generally a new state. 

Remarkably, it seems that this sort of thing does occur, at least
theoretically, in another class of candidate quantum Hall states
\cite{moore}.  These are states that in a sense combine the characteristic
Laughlin ordering with a form of BCS pairing.  The prototype state of
this type is the so-called `Pfaffian' state, which occurs at filling
fraction $\nu = 1/2$ (note the odd denominator, which cannot occur for
a conventional hierarchical state).  In this system, $2n$ quasiholes
fill out an irreducible $2^{n-1}$ dimensional representation of the
braid group.  Notice that the non-trivial structure begins only when
there are at least four of these particles.

I think it is fair to say that the mathematical and physical
understanding of these non-abelian statistics is still at a primitive
stage.  The dimension of the representations can be inferred from
convincing but indirect and heuristic arguments using conformal field
theory or by a rather painful explicit construction.  (As I write
this, a more tractable version of the conformal field theory is being
developed\cite{fradkin}.)  The braiding properties have been worked out
explicitly for four particles.  There are strong indications that when
more particles are involved they naturally fall into a spinor
representation in Hilbert space.  This property is eerily reminiscent
of D-particle quantum mechanics, where the statistics become promoted
to a continuous group.

Experimentally, neither the unusual abelian nor the nonabelian
statistics has yet been tested directly.  Indeed it is only quite
recently that the fractional charge of Hall effect quasiparticles was
convincingly observed, in shot noise experiments\cite{piccio}.  The
situation is far from hopeless, however.

In higher dimensions, the braid group is no longer relevant -- one has
room to untangle the braids -- and one is dealing simply with the
symmetric group.  This group of course supports non-abelian
representations, leading to the concept of parastatistics.  There is
another possibility, however, that is (in contrast to parastatistics)
remarkably unique and extends smoothly to arbitrary numbers of
particles.  This is the possibility of projective statistics.  As I
shall now discuss, it also leads to spinor representations closely
related to what arises in the Pfaffian state.

It is a general feature of quantum mechanics that symmetries of the
observables are implemented by unitary transformations of Hilbert
space.  However, this implementation need not respect the group
multiplication law, when there is a group of symmetries.  Since the
observables are not affected by a phase redefinition of the states,
all that is required is $U(g_1) U(g_2) = \eta (g_1, g_2) U(g_1 g_2)$,
in an obvious notation; the $\eta$s are complex numbers of modulus unity. 

The projective representations of the symmetric group are related to
ordinary representations of an extended group I shall write as $\tilde
S_N$, where $N$ is the number of identical particles.  $\tilde S_N$
plays, in this regard, the same role for the symmetric group $S_N$ as $SU(2)$
plays for the rotation group $SO(3)$.  
$\tilde S_N$ is most easily defined using generators and relations.
One introduces elements $T_i$, $1\leq  i\leq N-1$, and $z$ satisfying 
$T_i^2 = 1$, $(T_i T_{i+1} )^3 = 1$, and $T_i T_j = z T_j T_i$ for
$|i-j| \geq2$.  $z$ is a central element with $z^2 = 1$.  One should
think of $T_i$ as implementing the transposition $(i, i+1)$.  If $z=1$
these relations define the symmetric group itself; so the irreducible
representations of $\tilde S_N$ with $z = 1$ simply reproduce the
ordinary irreducible representations of $S_N$.  The other possibility
is $z= -1$.  This corresponds to projective representations
proper.   From the linear representation of $\tilde S_N$ one obtains a
projective representation of $S_N$ by choosing, for every permutation
in $S_N$, some definite way of obtaining it as a product of the transpositions 
$T_i$.  The choice is unique up to a sign ambiguity,
which cannot be removed.  A very curious aspect of the properly projective
representations is that the {\it order\/} of interchanges, even
between distinct and possibly distant particles, affects the phase of
the wave function; this is a direct consequence of the third defining
relation for $\tilde S_N$.  

The projective representations of $S_N$, or equivalently the linear
representations of $\tilde S_N$, were determined by Schur in a series
of remarkable papers, even before the advent of quantum mechanics
\cite{schur,hamer}.  The properly projective representations, like
the ordinary representations, are associated with Young tableaux.
There is the new feature, however, that the length of the rows must be
strictly decreasing.  Thus there is a projective analog of the bosonic
tableaux (one row of length $N$) but not of the fermionic tableaux
($N$ rows of length one).

The
simplest, quasi-bosonic projective representation is already quite richly 
structured, and naturally involves spinors in $N$ dimensions.  
One can realize it, although not quite irreducibly, using
a set of $N +1$ gamma matrices $\Gamma_i$, and defining $T_i = {1\over
\sqrt 2} (\Gamma_i - \Gamma_{i+1})$.  By a change of basis we can make
do with $N$ matrices, defining $T_i = \sqrt{i-1\over 2i} \Gamma_{i-1}
- \sqrt{i+1 \over 2i} \Gamma_i$.  In this form, the representation is
irreducible.   Its size grows exponentially with the number of
particles, as $2^{[N/2] - 1}$, just matching the corresponding
degeneracy of quasiholes in the Pfaffian state.  This sort of growth
can be interpreted as the emergence of a new two-valued degree of
freedom with each added {\it pair\/} of particles.  Such a picture can be made
more explicit by representing the $\Gamma$ matrices in terms of
fermion creation and annihilation operators, as 
$\Gamma_{2i-1} = a_i + a^*_i$, $\Gamma_{2i} = i(a_i - a^*_i)$.   

In closing I would like to add one more general observation on this
topic.  In the
generic quantum theory, the observables are independent of an overall
phase rotation, and this opens the possibility of projective
realizations of symmetry.  In gauge invariant systems the observables
are independent of local gauge transformations; in principle, this
opens up the possibility of much wider classes of `improper'
realizations of quantum symmetry.

\end{document}